# High-resolution human mobility data reveal race and wealth disparities in disaster evacuation patterns

Hengfang Deng[1], Daniel P. Aldrich[2], Michael M. Danziger[3], Jianxi Gao[4], Nolan E. Phillips[5,6], Sean P. Cornelius[7,*], Qi Ryan Wang[1,*]

[1] Department of Civil & Environmental Engineering, Northeastern University, Boston, MA, 02115, USA
[2] Department of Political Science, Northeastern University, Boston, MA, 02115, USA
[3] Network Science Institute, Northeastern University, Boston, MA, 02115, USA
[4] Department of Computer Science, Rensselaer Polytechnic Institute, Troy, NY, 12180, USA
[5] Department of Sociology, Harvard University, Cambridge, MA, 02138, USA
[6] Accenture, Arlington, VA, 22209, USA
[7] Department of Physics, Ryerson University, Toronto, ON, M5B 2K3, Canada
* To whom correspondence should be addressed

**Major disasters such as extreme weather events can magnify and exacerbate pre-existing social disparities, with disadvantaged populations bearing disproportionate costs. Despite the implications for equity and emergency planning, we lack a quantitative understanding of how these social fault lines translate to different behaviors in large-scale emergency contexts. Here we investigate this problem in the context of Hurricane Harvey, using over 30 million anonymized GPS records from over 150,000 opted-in users in the Greater Houston Area to quantify patterns of disaster-inflicted relocation activities before, during, and after the shock. We show that evacuation distance is highly homogenous across individuals from different types of neighborhoods classified by race and wealth, obeying a truncated power-law distribution. Yet here the similarities end: we find that both race and wealth strongly impact evacuation patterns, with disadvantaged minority populations less likely to evacuate than wealthier white residents. Finally, there are considerable discrepancies in terms of departure and return times by race and wealth, with strong social cohesion among evacuees from advantaged neighborhoods in their destination choices. These empirical findings bring new insights into mobility and evacuations, providing policy recommendations for residents, decision makers, and disaster managers alike.**

Natural hazards continue to threaten lives and property around the world, killing nearly 12,000 people and causing $130 billion US damage in 2019 alone[1] Societies have set up a variety of ways to mitigate extreme weather events such as flooding; the most effective involves removing residents from vulnerable areas in advance of the hazard. Evacuations are not without challenges, however: effective departure may be impeded by severe traffic, city-level dwindling resources (e.g., fuel shortage, etc.), and uninhabitable shelter conditions[2–4]. Understanding the

underlying patterns of collective, abrupt displacement is therefore crucial for emergency management agencies to execute comprehensive, well-ordered plans[5].

There has been an increasing interest in linking social vulnerability to disasters through frameworks such as equity and environmental justice[6]. Previous studies suggest that minority populations from disadvantaged neighborhoods are more likely to be affected by disasters[7]. Examinations of Hurricane Katrina, for example, showed that black and poor residents of New Orleans suffered disproportionately severe impacts[8]. Further, race and class gaps are present from the preparation to response phases of shocks as well as through the recovery period[9,10]. Much research has argued that White communities are better prepared for disasters than communities of color and that White residents return at higher rates than Black or Hispanic residents[11,12]. Another study using post-disaster survey data of residents affected by 2012 tornadoes shows that households with more advantaged characteristics such as social capital and social ties recover faster from the disaster[13]. Despite these critical findings, there is still a lack of large-scale quantitative research on the impact of wealth and racial difference on disaster evacuation and return. To this end, scholars have typically relied on retrospective surveys and interviews[14,15] which face challenges such as sample bias, faulty retrospective recall, and low response rates[16,17].

Recently, GPS-enabled devices which generate unprecedented amounts of mobility data have enabled more nuanced studies of behavior[18–21], including in large-scale disasters and extreme events[22,23], allowing the construction of probabilistic models[24,25], and highlighting the importance of social networks[26–28] and socio-economic factors[10]. By comparing multiple disasters, certain typical behaviors have been established, including an exponential return rate[29].

Hurricane Harvey made landfall on August 25th, 2017 and to-date is rivaled only by Hurricane Katrina as the costliest tropical cyclone on record. It was accompanied by wind speeds as high as 130 mph and more than 50 inches of rain, making it the wettest tropical cyclone in history and leaving no part of the city unscathed. Yet despite this hurricane's magnitude, authorities did not order a mandatory evacuation in Houston, providing a rare example of widespread evacuation that occurred emergently through countless individual decisions, without the intervention of the state.

Here, we take Hurricane Harvey as a case study, and use anonymized mobility data to identify detailed evacuation behavior. We then link the observed behavior with information on wealth and race at the level of the census block group—the smallest spatial unit of reporting of the US Census. This allows us to quantify the extent to which these socioeconomic factors impacted evacuation behavior. We show how these disparities influenced not only who evacuated, but also where they evacuated to, when they evacuated and how long they remained evacuated.

Houston itself is an ideal testbed to study the disparities of interest, being the fourth most populous city in the US and hosting a racially and economically diverse population.

We use anonymized data generated by more than 2 million opted-in mobile devices between July 1, 2017 to October 1, 2017 to illuminate the class and race interactions with disaster evacuation and recovery. More details about the data can be found in *SI Section 1, Data Description.* We present a multi-scale study of disaster-induced evacuations of Hurricane Harvey looking at who left, where they went, and how long they were gone. First, we infer residents' evacuation status and destination using spatial data analysis. We then examine the socio-demographic differences across the Houston metropolitan statistical area and integrate the information with the distributions of spatial and temporal patterns at the individual level. Finally, we use residential home neighborhood characteristics capturing both race and wealth to illuminate the differences in disaster responses across all affected regions.

## Results

**Who evacuated?**
When Hurricane Harvey approached the Texas coastline, officials did not instruct the populace in Houston to evacuate. And yet, a significant percentage of the population chose to evacuate anyway. Without fine-grained data, it is difficult to determine to what extent there were patterns of evacuation based on location or socioeconomic features. Based on our analysis of the mobility data, we were able to detect 10,179 unique evacuees and 141,828 non-evacuees, indicating that 6.7 percent of the total population evacuated, on par with the official reports. However, as we show in Fig. 1, this baseline evacuation rate varied substantially across different locations and neighborhood types. In addition to a higher evacuation rate closer to the coast, we also identify many inland areas with evacuation levels as high as the coastal areas, such as Fort Bend and East Houston areas (Fig. 1a). We further identify two distinct geographic patterns for long-distance ($90^{th}$ percentile, >41.25 km) and short-distance ($10^{th}$ percentile, <2.71 km) evacuations, respectively. Whereas long-distance evacuation was almost entirely directed from the coastline inland, short-distance evacuation occurred in distinct, concentrated pockets across the MSA.

We studied social disparity by classifying six types of neighborhoods. A neighborhood in this study is a census block group, a geographical unit used by the United States Census Bureau. Each block group contains between 600 and 3,000 residents. We divided the census block groups in the Greater Houston area into six classes based on the wealth level (using a threshold of 25% of the residents below the federal poverty level) and the majority racial group (with a threshold of 50%): non-poor Black, non-poor Hispanic, non-poor White and poor Black, poor Hispanic, and poor White.  We also conducted a sensitivity analysis by setting different thresholds on the poverty level and the majority racial group; the results can be found in *SI Section 3 Sensitivity Analysis*.

To investigate both geographical and sociodemographic features of the evacuation mobility flow induced by Hurricane Harvey, we start with a multivariate kernel density estimate with Gaussian kernels on both origins and destinations (see *Materials and Methods*) to produce the contour plot of net flow density in each 100 m x 100 m grid. We then compare all evacuees' residential block group characteristics (see Table 1) with the population breakdown to show the percentage difference as the relative evacuation rate.

| Type | Number of Block Groups | Total Population (ACS 2018) | Total Devices Identified | Population Baseline Ratio | Device Baseline Ratio |
|---|---|---|---|---|---|
| Non-Poor Black | 261 | 551,779 | 11,029 | 0.112 | 0.076 |
| Non-Poor Hispanic | 624 | 1,282,879 | 27,192 | 0.261 | 0.188 |
| Non-Poor White | 1064 | 2,278,045 | 93,483 | 0.474 | 0.647 |
| Poor Black | 108 | 166,702 | 3,539 | 0.034 | 0.018 |
| Poor Hispanic | 313 | 598,496 | 9,016 | 0.122 | 0.062 |
| Poor White | 26 | 36,558 | 1,213 | 0.007 | 0.008 |

**Table 1.** Statistics of each type of neighborhoods included in this study (Total Number of Block Groups, Total Population and the Total Number of Active Devices) and the corresponding population percentage as the baseline ratios.

Based on the locations from which the individuals evacuated, we can examine whether socioeconomic factors impacted evacuation behavior. After reweighting the mobility records to account for sampling bias (as described in *Materials and Methods*), we find that evacuees from non-poor majority White block groups are significantly overrepresented among evacuees, 19.8% more than we would expect based on their fraction of the total population (Fig. 1d). All other neighborhood types evacuated less than the baseline, even though flooding was comparable across the neighborhoods. On the opposite end of the spectrum, individuals from poor Hispanic communities showed the least tendency to evacuate with 12.2% below their baseline (Fig. 1d). The result aligns with observations from other natural disasters. For example, wealth has also been shown to correlate positively with evacuation following Hurricane Irma[10].

Through Fig. 1(b-c) and 1(e-f) we compare the subgroup patterns based on the evacuation distances. Fig. 1b and 1e show the individuals who had long-distance evacuations (>41.25 km), i.e., moved greater than the 90th percentile of the evacuation distance distribution. Fig 1b shows the spatial density flow corresponds to the evacuees who travelled long distances, where we can observe clear hotspots along the coast moving toward the inland areas. The patterns of evacuation rates from different neighborhoods remain largely the same as the ones observed in overall evacuees. The only difference is that we observe a slight increase (4.2%) of the individuals from non-poor White communities when comparing it to the overall relative evacuation rates (see Fig. 1e). Similarly, the evacuation discrepancy holds for those who relocated less than the 10th percentile (<2.71 km) even though the origins and destinations of evacuations are more spatially dispersed (Fig. 1c). However, nonpoor White neighborhoods are the only groups that provide more evacuees than their baselines (Fig.1f).

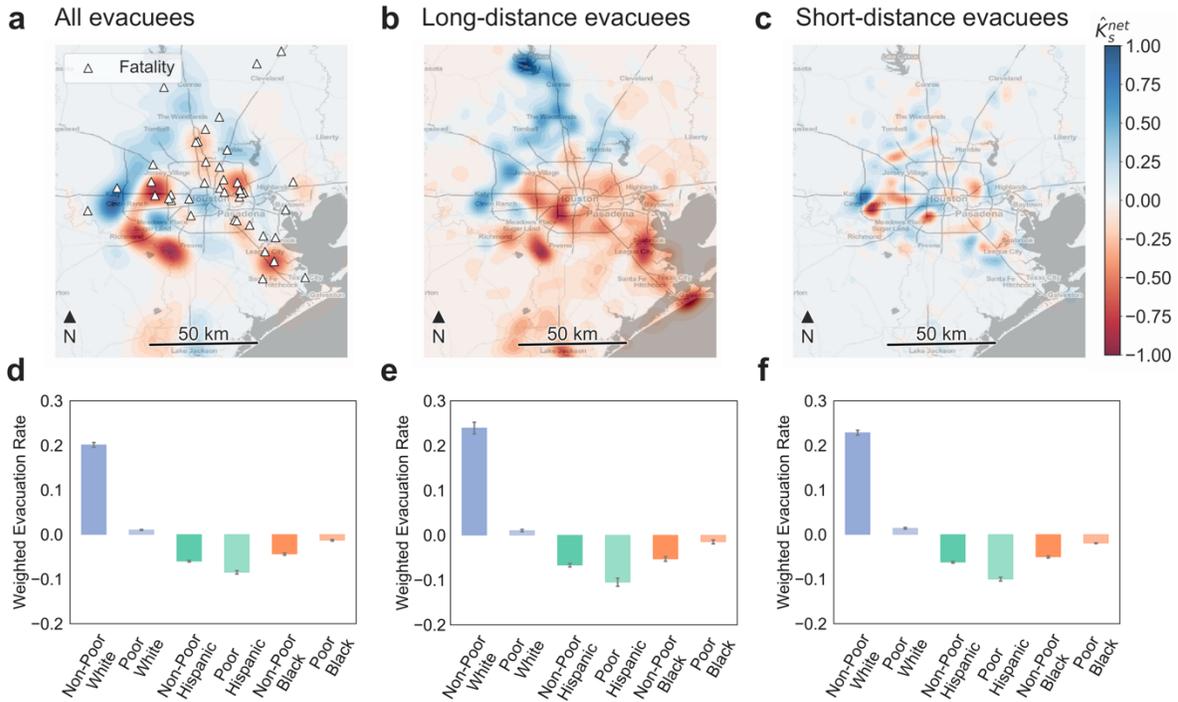

**Fig 1. Geographical and socio-demographic information on evacuation from Hurricane Harvey**. *a-c* illustrate the net *evacuation intensity* of each 100x100m grid, defined as the difference $\widehat{K}_s^{net} = \widehat{K}_s^{post} - \widehat{K}_s^{pre}$, where $\widehat{K}_s^{post}$ ($\widehat{K}_s^{pre}$) are KDE estimates of the probability density of home locations after (before) evacuation. These values are normalized to have zero mean and lie in the range [-1,1] (see Materials and Methods). As such, positive (blue) values correspond to an influx of resident above the baseline, with negative (red) values indicating an exodus. We compare the collective evacuation behavior of: all evacuees (*a*); evacuees who evacuated further than the 90% quantile (*b*); and evacuees who evacuated closer than the 10% quantile (*c*). The triangles in (*a*) represent reported fatalities and red dense cells indicate a greater rate of evacuation while the blue shaded cells reflect the likelihood as a relocation destination for the evacuees. The relative evacuation rates that compare the proportion difference for subgroups of populations are shown in (*d*) though (*f*) and an evident disparity in the evacuees' economic and social compositions can be observed.

### Where did they go?

In addition to detecting who left vulnerable areas during shocks, it is important to understand the locations to which they evacuated. This understanding is important for disaster managers who seek to efficiently pre-position shelters with food, water, and medical care and also for police officials and first responders who seek to ameliorate transportation challenges, such as gridlock. Without fine-grained mobility data, these remain major policy challenges. In Fig. 1 (a-c), we can see the general distribution of evacuation sources and destinations, but the evacuation distance is not discernable. We find that, despite the wide differences in evacuation rates for different neighborhood types discussed above, the evacuation distance follows a scale-free distribution with an exponential cutoff, in line with findings from other disasters**Error! Bookmark not defined.**. Specifically, the distribution $P(\Delta d) = (\Delta d_0 + \Delta d)^{-\beta} exp(-\Delta d/K)$ fits for evacuees from different communities with exponent value $\beta = 1.57$, $\Delta d_0 = 2.19$ km and cutoff value $K = 38.29$ km.

Based on this apparent universality, we might expect that the evacuation process is spatially uniform, with evacuation destinations determined only by one's distance from their origin. We find that this is not in fact the case; there is remarkable socioeconomic homophily in evacuation behavior, with people from a given block group type tending to evacuate to the same block group type, even if it is far away. To assess this, we compute the transition matrix between block group types based on the origin-destination information for all evacuees. We normalize the matrix so that the values represent the evacuees' fraction from a block group type who evacuated to a given block group type. In Fig. 2b and c we see strikingly different behavior from the spatial pattern. The residents from White neighborhoods have an 88.1% probability of evacuating to the same type of block group, while for Hispanic communities that ratio dropped to 56.8%. Evacuees from Black communities are the least likely to relocate to a neighborhood that is consistent with their original home community type (16.7%). Similarly, individuals from non-poor communities have a 92.9% probability to relocate to a similar social background. In contrast, the rate drops to 35.x% for residents of poor neighbourhoods..

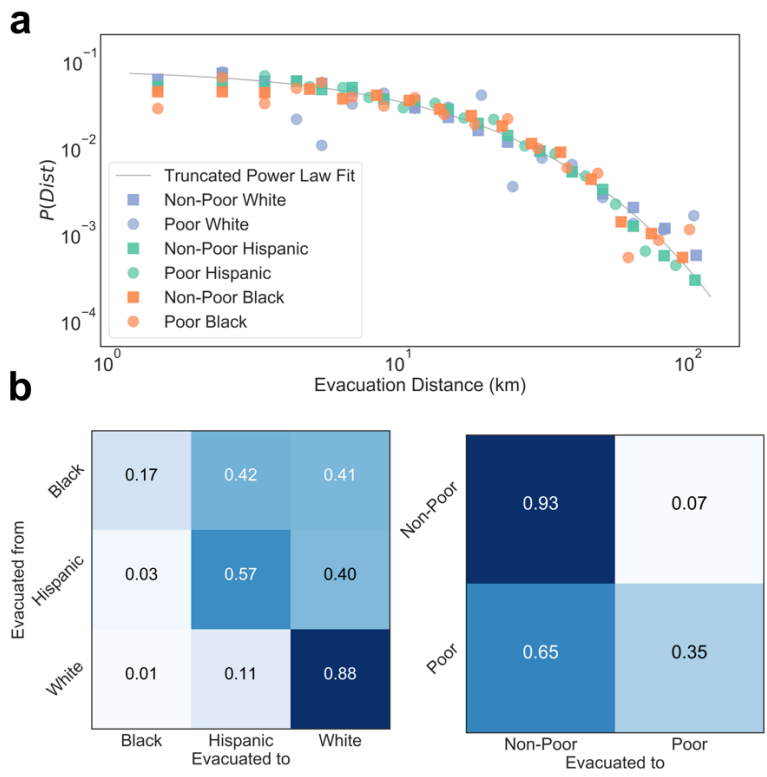

**Fig 2. Evacuation distance and destination choice. (a)** The distribution of evacuation distance for residents from different block group categories; **(b)** the correspondence between evacuation origin and destination type (note: all rows sum to 1). There is apparent universality between different block group types when viewed from the probability of evacuating a given distance **(a)**. But this apparent similarity breaks down upon examining destination choices, with residents of wealthier **(b, right)** and/or white neighbourhoods tending to evacuate to block groups of the same composition.

**How long were they gone?**
Evacuation duration is a key factor in disaster recovery and resilience[16,30,31]. Identifying the precise beginning and end of evacuation using the mobility data, we find that most evacuation departures occurred in the first few days following landfall, with 95.6% between 1 and 7 days following landfall. Only a few people (4.3%) evacuated in advance of landfall. In contrast, returns began shortly after the peak of departures but spread out over a more skewed, heavier-tailed distribution; though most evacuees returned within two weeks, some took far longer. The exact dates of departure and return were determined via spatial clustering with a 3-day sliding window (see *SI Section 2*) to reflect whether an individual stabilized in a given location.

As with the distance distribution, it is tempting to think that the approximately exponential return time distribution indicates a homogeneous behavioral pattern. Yet when we incorporate the sociodemographic information of each evacuee, we uncover marked socio-economic disparities. We perform the bootstrapping on the overall samples iteratively for the different neighborhood types compared to the baselines based on the cumulative evacuees and returners as a subgroup. From the violin plot of evacuation duration shown in Fig. 3b, we can see that, as with the spatial patterns, the basic statistics such as mean/median show little variation across different types of block groups. 90% of poor Black evacuees returned after 13 days, whereas at that point only 75% of poor White evacuees had returned, indicating huge variability in return times. Additionally, the vast majority of long-duration relocations (over 30 days) involve evacuees from higher social class neighborhoods *(see SI Section 4)*. Past research has suggested that poorer residents have greater challenges finding new, short term housing while wealthier and better positioned ones can do so more easily[32].

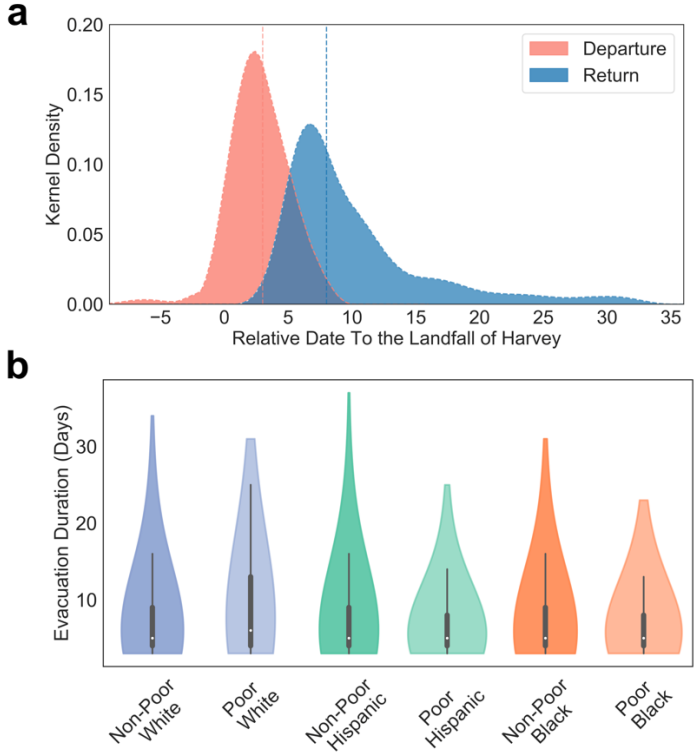

**Fig 3. Race and wealth impacts on evacuation/return times**. **(a)** The kernel density of both departure and return of relocation activities. **(b)** Distribution of individual evacuation durations group by income and race difference in communities. Though most departure dates are tightly clustered within the first week of Harvey's landfall **(a, red)**, we find a right-skewed distribution of return dates **(a, blue)**, with some Houston residents taking far longer to return than others. The tail of this distribution is form mostly of residents of wealthier neighbourhoods of all racial compositions **(b)**, suggesting only those residents could afford long displacements.

We are further interested in discovering whether socio-demographic factors impact the time of evacuation day by day, for example, to determine whether certain groups are more likely to have evacuated immediately upon landfall or to have waited. A similar question regards return from evacuation: on a given day, which groups are over or underrepresented? To address these questions, we compute the disparity rate in departure and return times. Due to the limited number data points in certain block group types, we calculate two disparity rates based on two aggregations of neighbourhoods into i) non-poor vs. poor; and ii) white majority vs. non-white majority, as defined above. For each such division, we then find the relative difference in the cumulative number of prevalence of departures/returns on each day. The racial (wealth) disparity rate $D_i$ for a given day $i$ is thus calculated as $(R_i - R)/R$, where $R$ is the total fraction of evacuees coming from white (non-poor) neighbourhoods while $R_i$ is the same fraction, but only including evacuations/departures up to day $i$. High values of $D_i$ thus indicate that evacuations (or returns) from white or non-poor neighbourhoods were overrepresented up to time $i$ in the process.

Fig. 4a indicates that while most departures took place within a short period, the social disparities persist along the temporal dimension. In terms of poverty (or wealth), we see that for those who departed prior to the disaster, the percentage of residents from wealthy neighborhoods is over 50% higher than that of all evacuees. Such variation presents a steady decrease as the hurricane strikes the city and falls below 50% after the first day of the landfall. We also observe that the standard deviation is notably larger in the early days which is due to the small number of early evacuees.

Overall the disparity rates along the wealth dimension were positive over multiple days, suggesting that the people with higher income were more likely to evacuate ahead of those with lower wages. The trend based on race also shows that evacuees from neighborhoods with higher White-population ratios were more likely to evacuate early, yet right after the landfall people from disadvantaged block groups comprised more of the evacuees (with about 10% higher than the baseline). Overall, both income and race disparity metrics indicate that majorities of early evacuees comprise people from potential higher social classes, yet this disparity decreases at a steady pace.

When we project the disproportionate effect of return behavior over the period between 28 August and 17 September which is shown in Fig. 4b, we observe that, like the departure behavior, the returners' daily disparities show greater class than racial difference. We consider Day 5 after Harvey as the beginning of the return period based on local news and weather reports verifying rainfall had stopped and Day 23 as the end, based on the 95th quantile of the overall return time. We see that early returners, especially those who came back to their homes in less than 3 days, primarily came from the wealthy communities. However, this class disparity effect continues to decline with a greater proportion of poor evacuees returning and reaches equity 2 weeks after the disaster. Such class differences increase monotonically, reaching the same proportion after 3 weeks, at which point most of the evacuees have returned to their homes in Houston.

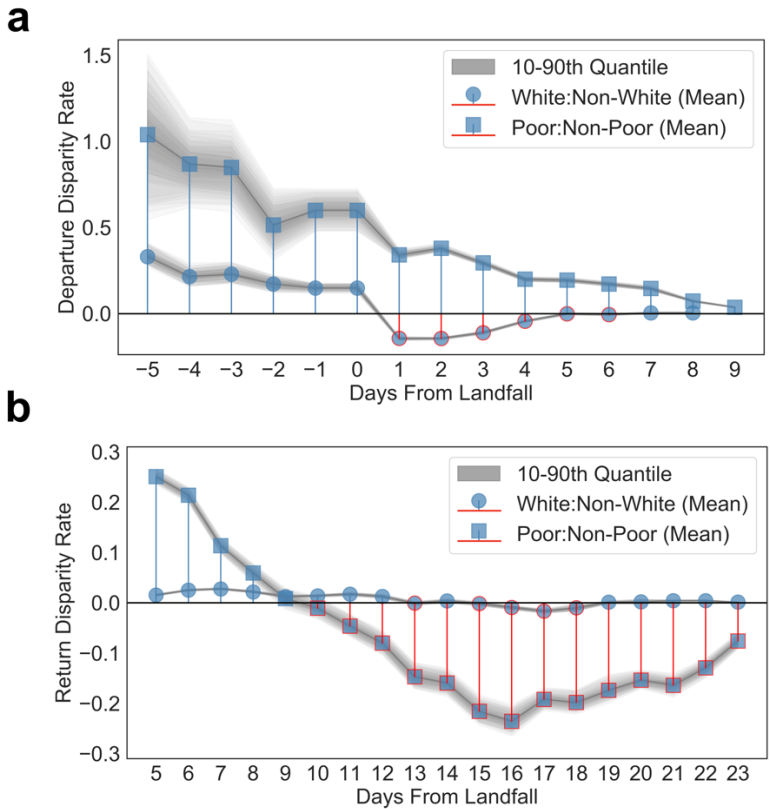

**Fig 4. Time progression of class and racial disparity in evacuation and return.** The vertical axis shows the relative disparity in the cumulative numbers of departures **(a)** and returns **(b)** over time, between residents of poor vs. non-poor neighbourhoods (squares), and white vs. non-white neighbourhoods (circles). A positive value on a given day indicates a disproportionate number of people from white or non-poor neighbourhoods having evacuated (returned) up to that point.

Finally, we estimate the extent to which the evacuation and recovery timing varied by race and class by computing the daily disparity rate. We find that a small percentage of the overall evacuees (4.3%) responded to Harvey and relocated before its landfall. The low rate of evacuation may be because urban residents collectively perceived a low level of risk in the absence of mandatory evacuation orders in Houston. Particularly among early evacuees there are marked disparities, suggesting that the ratio of the wealthy residents has a 50% jump compared to the evacuee baseline and 25% for the ratio of people residing in White majority neighborhoods. As expected, the class and racial minority proportion of the evacuees increases steadily after Harvey's arrival. This finding confirms that minorities are at higher risks of disruption and adverse impacts. This behavioral difference indicates that income level could be a contributing factor in early-evacuation decision-making. Further, disparities in terms of time away indicate that income plays an important role. We can see that the proportion of evacuees from disadvantaged communities started increasing as the storm ended, echoing previous studies[10,16] that class and housing ownership has a strong impact on return dates.

# Discussion

Our analysis of three months of detailed human mobility data before, during, and after Hurricane Harvey reveals the complexity of disaster response patterns, exhibiting both universality and heterogeneity in different dimensions. We first quantify the wealth and racial disparity in terms of the overall evacuee composition as well as the intra-urban displacement patterns. By examining these hotspots, we find that three of the peaks correspond to areas with high fatalities while the other hotspots to the southwest of downtown Houston had no fatalities. We further observe that non-poor White populations were the most likely to evacuate, for evacuations of all distances. There are several plausible reasons for such class and race differences, including uneven access to transportation, inequitable regional development and perhaps even different levels of preparedness and perceived risks. In Harvey's case, with no evacuation order declared for Houston, the evacuation decision process may have taken place with little prior planning, increasing the role of financial resources and social networks in making such decisions.

Furthermore, we extract the spatial and temporal features of evacuation activities and investigate the heterogeneity among different residential groups for each metric. There is a high degree of universality in evacuation distance after comparing the evacuees' distributions from different types of communities. This finding aligns with previous studies based on similar extreme disasters such as hurricanes and earthquakes, indicating that evacuation during Harvey followed the heavy-tailed human movement pattern of past crises. The overall temporal features demonstrate limited impact from differences in the socio-demographic features of communities. For most of the block groups, the evacuees' displacement duration was less than one week while residents from more affluent communities were displaced for longer. It is also notable that displacements longer than 30 days are disproportionately residents of socio-economically advantaged neighbourhoods. The longer exodus is likely because only the wealthy possess the financial resources to afford a longer stay away; those with low incomes would struggle with limited resources and potentially higher burdens.

Even with the high spatial and temporal resolution of our data, our analysis is not without limitations. First, we have focused only on mobility in Houston MSA during a three-month period between July and October 2017. Larger areas and longer observation windows would be required to detect larger (e.g., out of state) displacements and long-term impacts such as permanent relocations. A second limitation is the sampling bias amongst different communities owing to different penetration rates of smart devices. Though we have made efforts to quantify and mitigate this bias through resampling (see *Material and Methods*, *Sampling Bias Correction*), ensuring data balance remains a challenge. Third and finally, our analysis of race and class has not fully controlled for other features known to impact evacuation behavior, such as home damage, infrastructure conditions, and strength of social ties. We have tested a regression model incorporating data on flooded roads to capture infrastructure conditions and found the overall predictive power of that factor is not significant (see *SI Section 5, Evacuation Rate Prediction*). Nonetheless, additional information about the urban fabric—encompassing both physical and social infrastructure—is needed to fully understand the complex interplay between race, wealth, and disaster outcomes.

The significance of income levels in determining evacuation response, with wealthier residents evacuating earlier, longer, and at a higher overall rate, indicates that some residents may have preferred to evacuate but judged that the costs of evacuation were too high, and they therefore took on the risks of staying put. This impossible choice between physical harm and economic harm is a recurring theme, notably in the very different crisis of COVID-19. In the first wave of spring 2020, compliance with social distancing guidelines was significantly lower in low-income neighborhoods across the United States[33,34]. This points to potential interventions in the form of temporary financial incentives, government messaging, and/or preferential allocation of resources aimed at ameliorating the unequal effects of major urban disruptions.

# Material and Methods

**Data and Preprocessing**
We use mobility data provided by Cuebiq (www.cuebiq.com), a leading private sector firm that handles human mobility data with anonymized identifiers and guarantees user privacy. Prior studies have used these and similar data sets to understand commuting patterns[35], accessibility to public resources[36], neighborhood connectedness[37], and social distancing during epidemics[38,39]. We confine the geographical bounds to the Houston Metropolitan Statistical Area (MSA) and collect the de-identified mobility data for a three-month period overlapping with Hurricane Harvey (July 1st to September 30th, 2017), comprising over 3 billion geographical records from over 2.5 million unique anonymous users who opted-in to share their location data anonymously for research purposes. Due to differences in hardware/software among the devices in the dataset, the GPS coordinates therein possess different levels of spatial noise. We adopt a uniform minimum spatial granularity of 50 m by implementing an established stay-point algorithm[40,41] to extract meaningful stays and filter out transient locations. The algorithm works by processing a de-identified GPS trajectory (timestamped sequence of latitude/longitude pairs) into a series of geographic regions where the user has a stay duration surpassing a specified threshold. The effect is to compress trajectories, allowing efficient analysis of individual mobility patterns; each anonymized device's trajectory becomes a series of stay points encoded as tuples of the form (latitude, longitude, start_time, end_time). Throughout this study, we use temporal and spatial thresholds of 5 minutes and 50 meters, respectively. To minimize the impact of tourists and other non-residents, we retain only users who have at least data points from at least 60 unique days and at least 100 stay points. This filtering leaves roughly 30 million data points corresponding to roughly 150,000 unique, anonymous users. The detailed data specifications are shown in SI Table 1.

**Sampling Bias Correction**

A potential challenge with crowdsourced large-scale mobility data is the inherent bias in sampling rates for different populations. To test how this may confound the results that follow, for each block group we compare the number of devices in the dataset with the total population reported in the 2018 American Community Survey (ACS 2018), as detailed in *SI Section* 1.3 Data Representativeness. We observe strong correlation between the two datasets, although the distributions do not align perfectly.

To minimize potential bias, we employ two strategies: i) a weighting procedure, in which we weight each block group according to the ratio of number of actively reporting devices to the reported population based on ACS 2018. By comparing both the unweighted and weighted results, we can gauge the magnitude of deviation of the results caused by the sampling bias. However, in the case when the subgroup of evacuees was small or where evacuees from a small number of specific neighborhoods predominate, the sampling bias from those neighborhoods can appear to be more prominent than the neighborhood disparity. Therefore, we also ii) bootstrap the samples from each block group with uniform sampling rate proportional to the population with 100 iterations.

**Home Census Block Group Detection**

For every anonymized device in the dataset, we first determine the home locations prior to the landfall of the disaster by identifying the weekly primary locations using stay points overlapping the weekday evenings (8 pm to 7 am). We then apply an agglomerative clustering method using complete linkage[42] (enforcing a maximum spatial diameter threshold of 50 m to each cluster) and assign the cluster region with the longest cumulative stay duration as the candidate home census block groups. In addition, a threshold of minimum two different days has been applied to mitigate the uncertainty of unbalanced usage across different evenings. After obtaining 5 consecutive weeks of potential home census block groups for each individual, we perform a weekly cross-validation with tolerance of only one week missing or mismatch, using 50 m as the maximum deviation Haversine distance in consistency of the stay-point accuracy.

**Evacuation Detection**

To determine each user's evacuation status and identify the corresponding evacuation and return time, we use the cross-validated home location (above) for each user as the prime baseline location. The key challenges are the uncertainty for daily reported data as well as the uncertainty of mislabeling short trips as evacuations. We therefore introduce a sliding time window (testing both 5 and 7 days) to mitigate such risks and iteratively identify the primary home locations of each sliding time window, computing the deviation distance to the prime location. Within each rolling time window, we apply the same home detection technique and determine primary location. If a departure of at least three consecutive calendar days from home is observed, we consider this user as a potential evacuee and label the new primary location as the evacuation destination. We used a 1 km distance threshold of evacuation for the cluster maximal diameter from 50 m to, in order cope with the fact that the accuracy of signal and cellular service could be affected by Hurricane Harvey. The middle date of the time window is taken as the date of departure.

**Net Evacuation Intensity Estimation**

In order to estimate the geographical patterns of intra-urban relocation, we use non-parametric Kernel Density Estimation (KDE) to evaluate the net migration intensity for each urban grid $s$ denoted as $\widehat{K}_s^{net}$. Specifically, we use the pre-disaster and post-disaster latitudinal and longitudinal information of all evacuee's detected home locations as inputs for two KDE models that capture the probability density of home locations before and after evacuation, respectively. Each density function $\hat{f}$ has the following form:

$$\hat{f}(s) = \sum_{i=1}^{n} \frac{1}{nr^2} K\left(\frac{d_{i,s}}{r}\right)$$

where $s$ represents a specific location of interest and $r$ denotes the bandwidth of the KDE estimate, which relates to model resolution. In this study we and Gaussian kernel function $K$, with $r = 100$ m. Here, $i$ runs over all $n$ home stay points in the time period of interest, and $d_{i,s}$ is the Haversine (geodesic) distance between home location $i$ and the location of interest, $s$.

We use the set of original detected home location data of all evacuees and construct the spatial density function $\widehat{K}_s^{pre}$. Similarly, we can obtain $\widehat{K}_s^{post}$ with the post-evacuation destination. We then compute the *evacuation intensity* of each area as $\widehat{K}_s^{net} = \widehat{K}_s^{post} - \widehat{K}_s^{pre}$. In order to facilitate the comparison of contour surfaces based on different subgroups with varying sample size, the final values of $\widehat{K}_s^{net}$ are normalized to have zero mean and lie in the range [-1, 1]. This allows us to easily depict influxes (outfluxes) of people above/below the baseline.

# References

1. Guha-Sapir, D., Below, R. & Hoyois, P. EM-DAT: The CRED/OFDA International Disaster Database. *Centre for Research on the Epidemiology of Disasters (CRED), Université Catholique de Louvain* (2009).
2. Wolshon, B. & McArdle, B. Temporospatial analysis of Hurricane Katrina regional evacuation traffic patterns. *J. Infrastruct. Syst.* (2009) doi:10.1061/(ASCE)1076-0342(2009)15:1(12).
3. Dow, K. & Cutter, S. L. Emerging hurricane evacuation issues: Hurricane Floyd and South Carolina. *Natural Hazards Review* (2002) doi:10.1061/(ASCE)1527-6988(2002)3:1(12).
4. Wolshon, B., Hamilton, E. U., Levitan, M. & Wilmot, C. Review of policies and practices

for Hurricane evacuation. II: Traffic operations, management, and control. *Natural Hazards Review* (2005) doi:10.1061/(ASCE)1527-6988(2005)6:3(143).
5. Wang, J. W., Wang, H. F., Zhang, W. J., Ip, W. H. & Furuta, K. Evacuation planning based on the contraflow technique with consideration of evacuation priorities and traffic setup time. *IEEE Trans. Intell. Transp. Syst.* (2013) doi:10.1109/TITS.2012.2204402.
6. Adeola, F. O. & Picou, J. S. Hurricane Katrina-linked environmental injustice: race, class, and place differentials in attitudes. *Disasters* (2017) doi:10.1111/disa.12204.
7. Bolin, B. & Kurtz, L. C. Race, Class, Ethnicity, and Disaster Vulnerability. in (2018). doi:10.1007/978-3-319-63254-4_10.
8. Lovekamp, W. E. The Sociology of Katrina: Perspectives on a Modern Catastrophe. *Contemp. Sociol. A J. Rev.* (2008) doi:10.1177/009430610803700336.
9. Laska, S. & Morrow, B. H. Social vulnerabilities and Hurricane katrina: An unnatural disaster in new Orleans. *Mar. Technol. Soc. J.* (2006) doi:10.4031/002533206787353123.
10. Yabe, T. & Ukkusuri, S. V. Effects of income inequality on evacuation, reentry and segregation after disasters. *Transp. Res. Part D Transp. Environ.* (2020) doi:10.1016/j.trd.2020.102260.
11. Aldrich, D. P. & Meyer, M. A. Social Capital and Community Resilience. *Am. Behav. Sci.* (2015) doi:10.1177/0002764214550299.
12. Tierney, K. The social and community contexts of disaster. (1989).
13. Sadri, A. M. *et al.* The role of social capital, personal networks, and emergency responders in post-disaster recovery and resilience: a study of rural communities in Indiana. *Nat. Hazards* **90**, 1377–1406 (2018).
14. Reinhardt, G. Y. Race, Trust, and Return Migration. *Polit. Res. Q.* **68**, 350–362 (2015).
15. Houston, J. B. *et al.* Social media and disasters: A functional framework for social media use in disaster planning, response, and research. *Disasters* (2015) doi:10.1111/disa.12092.
16. Fussell, E., Sastry, N. & Vanlandingham, M. Race, socioeconomic status, and return migration to New Orleans after Hurricane Katrina. *Popul. Environ.* (2010) doi:10.1007/s11111-009-0092-2.
17. Rufat, S., Tate, E., Burton, C. G. & Maroof, A. S. Social vulnerability to floods: Review of case studies and implications for measurement. *Int. J. Disaster Risk Reduct.* (2015) doi:10.1016/j.ijdrr.2015.09.013.
18. González, M. C., Hidalgo, C. A. & Barabási, A. L. Understanding individual human mobility patterns. *Nature* (2008) doi:10.1038/nature06958.
19. Hasan, S., Schneider, C. M., Ukkusuri, S. V. & González, M. C. Spatiotemporal Patterns of Urban Human Mobility. *J. Stat. Phys.* (2013) doi:10.1007/s10955-012-0645-0.
20. Alessandretti, L., Sapiezynski, P., Lehmann, S. & Baronchelli, A. Multi-scale spatio-temporal analysis of human mobility. *PLoS One* **12**, e0171686 (2017).
21. Alessandretti, L., Sapiezynski, P., Sekara, V., Lehmann, S. & Baronchelli, A. Evidence for a conserved quantity in human mobility. *Nat. Hum. Behav.* **2**, 485–491 (2018).
22. Bagrow, J. P., Wang, D. & Barabási, A. L. Collective response of human populations to large-scale emergencies. *PLoS One* (2011) doi:10.1371/journal.pone.0017680.
23. Bengtsson, L., Lu, X., Thorson, A., Garfield, R. & von Schreeb, J. Improved Response to Disasters and Outbreaks by Tracking Population Movements with Mobile Phone Network Data: A Post-Earthquake Geospatial Study in Haiti. *PLoS Med.* **8**, e1001083 (2011).
24. Song, X. *et al.* Modeling and probabilistic reasoning of population evacuation during large-scale disaster. in *Proceedings of the ACM SIGKDD International Conference on*

*Knowledge Discovery and Data Mining* (2013). doi:10.1145/2487575.2488189.
25. Song, X., Zhang, Q., Sekimoto, Y. & Shibasaki, R. Prediction of human emergency behavior and their mobility following large-scale disaster. in *Proceedings of the ACM SIGKDD International Conference on Knowledge Discovery and Data Mining* (2014). doi:10.1145/2623330.2623628.
26. Yabe, T., Ukkusuri, S. V. & C. Rao, P. S. Mobile phone data reveals the importance of pre-disaster inter-city social ties for recovery after Hurricane Maria. *Appl. Netw. Sci.* (2019) doi:10.1007/s41109-019-0221-5.
27. Metaxa-Kakavouli, D., Maas, P. & Aldrich, D. P. How social ties influence hurricane evacuation behavior. *Proc. ACM Human-Computer Interact.* (2018) doi:10.1145/3274391.
28. Finch, C., Emrich, C. T. & Cutter, S. L. Disaster disparities and differential recovery in New Orleans. *Popul. Environ.* (2010) doi:10.1007/s11111-009-0099-8.
29. Yabe, T., Tsubouchi, K., Fujiwara, N., Sekimoto, Y. & Ukkusuri, S. V. Understanding post-disaster population recovery patterns. *J. R. Soc. Interface* **17**, (2020).
30. Sadri, A. M., Ukkusuri, S. V. & Gladwin, H. The Role of Social Networks and Information Sources on Hurricane Evacuation Decision Making. *Nat. Hazards Rev.* **18**, 04017005 (2017).
31. Gehlot, H., Sadri, A. M. & Ukkusuri, S. V. Joint modeling of evacuation departure and travel times in hurricanes. *Transportation (Amst).* **46**, 2419–2440 (2019).
32. SAMHSA. *How Disasters Affect People of Low Socioeconomic Status*. *Disaster Technical Assistance Center Supplemental Research Bulletin* https://www.samhsa.gov/sites/default/files/programs_campaigns/dtac/srb-low-ses.pdf (2017).
33. Ruiz-Euler, A., Privitera, F., Giuffrida, D., Lake, B. & Zara, I. Mobility Patterns and Income Distribution in Times of Crisis: U.S. Urban Centers During the COVID-19 Pandemic. *SSRN Electron. J.* (2020) doi:10.2139/ssrn.3572324.
34. Oliver, N. *et al.* Mobile phone data for informing public health actions across the COVID-19 pandemic life cycle. *Science Advances* vol. 6 eabc0764 (2020).
35. Wang, F., Wang, J., Cao, J., Chen, C. & Ban, X. (Jeff). Extracting trips from multi-sourced data for mobility pattern analysis: An app-based data example. *Transp. Res. Part C Emerg. Technol.* **105**, 183–202 (2019).
36. Akhavan, A. *et al.* Accessibility Inequality in Houston. *IEEE Sensors Lett.* (2018) doi:10.1109/lsens.2018.2882806.
37. Wang, Q., Phillips, N. E., Small, M. L. & Sampson, R. J. Urban mobility and neighborhood isolation in America's 50 largest cities. *Proc. Natl. Acad. Sci. U. S. A.* (2018) doi:10.1073/pnas.1802537115.
38. Gao, S., Rao, J., Kang, Y., Liang, Y. & Kruse, J. Mapping county-level mobility pattern changes in the United States in response to COVID-19. *SIGSPATIAL Spec.* **12**, 16–26 (2020).
39. Zhang, L. *et al.* an Interactive Covid-19 Mobility Impact and Social Distancing Analysis Platform. *Medrxiv* 1–14 (2020) doi:10.1101/2020.04.29.20085472.
40. Li, Q. *et al.* Mining user similarity based on location history. *GIS Proc. ACM Int. Symp. Adv. Geogr. Inf. Syst.* 298–307 (2008) doi:10.1145/1463434.1463477.
41. Jiang, S. *et al.* The TimeGeo modeling framework for urban motility without travel surveys. *Proceedings of the National Academy of Sciences of the United States of America* (2016) doi:10.1073/pnas.1524261113.

42. Murtagh, F. & Contreras, P. Algorithms for hierarchical clustering: An overview. *Wiley Interdiscip. Rev. Data Min. Knowl. Discov.* **2**, 86–97 (2012).

**Acknowledgments**: This work was supported by the National Science Foundation (NSF 1735505 and 1761950).

**Data availability:** Our data usage agreement with Cuebiq does not allow us to make public or otherwise share the anonymized mobile phone data used in this study. Researchers interested in aggregated data and/or summary statistics, where permitted under said agreement, should contact the corresponding authors.

**Code availability**: The code used to generate the results of this paper is available from the corresponding authors upon request.

**Ethics declaration (competing interests):** The authors declare no competing interests.